\documentclass[preprint,showpacs,preprintnumbers,amsmath,amssymb,floatfix,endfloats*]{revtex4}

\usepackage{graphicx}
\usepackage{dcolumn}
\usepackage{bm}
\usepackage{amsfonts}

\DeclareMathAlphabet{\mathsfsl}{OT1}{cmr}{bx}{it}
\begin{document}
\title{The effect of a reversible shear transformation on plastic deformation of an amorphous solid}
\author{Nikolai V. Priezjev}
\affiliation{Department of Mechanical and Materials Engineering,
Wright State University, Dayton, OH 45435}
\date{\today}
\begin{abstract}

Molecular dynamics simulations are performed to investigate the
plastic response of a model glass to a local shear transformation in
a quiescent system.      The deformation of the material is induced
by a spherical inclusion that is gradually strained into an
ellipsoid of the same volume and then reverted back into the sphere.
We show that the number of cage-breaking events increases with
increasing strain amplitude of the shear transformation.    The
results of numerical simulations indicate that the density of cage
jumps is larger in the cases of weak damping or slow shear
transformation.     Remarkably, we also found that, for a given
strain amplitude, the peak value of the density profiles is a
function of the ratio of the damping coefficient and the time scale
of the shear transformation.

\end{abstract}




\maketitle

\section{Introduction}

The mechanical properties of bulk metallic glasses, such as high
strength and low ductility, are both of fundamental scientific
interest and technological importance~\cite{Schuh07}.     It is now
well recognized that the plastic deformation of metallic glasses
below their glass transition temperature involves irreversible
rearrangements of small clusters of atoms~\cite{Argon79}.    The
plastic flow of amorphous materials in response to applied shear
stress can be described using the shear transformation zone model,
which takes into account the density and internal state of the
localized zones~\cite{Falk11}.   In recent years, various
deformation mechanisms including elementary plastic events and shear
band formation were studied at different length and time scales
using atomistic simulations and finite element
modeling~\cite{Rodney11}.   Notably, the energy landscape analysis
had shown that a large strain cycle rejuvenates the glass by
increasing the potential energy, while a small strain cycle overages
the glass by moving the system to deeper energy
minima~\cite{Lacks04}. However, many essential features of the
deformation process in strained amorphous systems including a
correlation between localized plastic events and distribution of
avalanches are not fully understood.

\vskip 0.05in

The effect of inertia on steadily sheared disordered solids in the
athermal quasistatic limit was examined in two and three dimensions
using molecular dynamics simulations~\cite{Salerno13}.      It was
found that the distribution of avalanche sizes obeys a power-law
decay over about three orders of magnitude in drops of the potential
energy density and shear stress, and the volume of plastically
deformed regions is proportional to the energy dissipated in an
avalanche~\cite{Salerno13}.       In the underdamped regime, the
system can be carried over successive energy barriers to
progressively lower minima leading to large avalanches, while in the
overdamped case, avalanches are smaller and they typically consist
of several disconnected regions oriented along diagonal
lines~\cite{Salerno13}.      It was also shown that at finite strain
rates and zero temperature, the correlation between local plastic
events remains relevant, and the avalanche size scales as the
inverse square root of strain rate in two
dimensions~\cite{Lemaitre09}.

\vskip 0.05in

In the last few years, a number of studies investigated oscillatory
shear response of amorphous materials using atomistic
simulations~\cite{Priezjev13,Sastry13,Reichhardt13,Priezjev14,Losert14},
continuum modeling~\cite{Perchikov14}, and experimental
measurements~\cite{Spaepen14,Cipelletti14,Arratia14,Echoes14,Arratia15}.
It was found that below a certain strain amplitude, the disordered
systems gradually evolve into dissipative limit cycles and particle
rearrangements remain reversible, thus retaining memory of the
initial
conditions~\cite{Sastry13,Priezjev13,Reichhardt13,Arratia14,Losert14}.
The number of back and forth cycles required to reach steady state
increases as the critical strain amplitude is approached from
below~\cite{Sastry13,Reichhardt13}.   Surprisingly, it was shown
that the cyclic deformation is accompanied by plastic rearrangements
of atoms that are reversed by the end of each
cycle~\cite{Reichhardt13,Arratia14}. With further increasing strain
amplitude, particle displacements become irreversible leading to a
diffusive behavior and structural
relaxation~\cite{Priezjev13,Sastry13,Priezjev14,Losert14}.

\vskip 0.05in

The elastic response of a two-dimensional amorphous solid to a
localized shear transformation was recently studied via molecular
dynamics simulations in different damping
regimes~\cite{Puosi14,Nicolas15}.       In this process, about
twenty atoms within a circular inclusion were instantaneously
sheared in a quiescent system and the time evolution of the
displacement field was measured.    It was demonstrated that the
stationary solution for the disorder-averaged displacement field has
a quadrupolar symmetry and it agrees well with the predictions of
the continuum elasticity theory~\cite{Puosi14,Eshelby57}.      It
was further observed that the transient regime is strongly dependent
on the damping dynamics and the time dependence of the displacement
field obtained from molecular dynamics simulations agrees with the
continuum solution in the overdamped case at large
times~\cite{Puosi14}.        The numerical analysis based on the
finite element method that takes into account microscopic viscosity
and the local elastic constants showed that the temporal evolution
of the disorder-averaged displacement field is similar to the
propagation of the elastic signal in a uniform
medium~\cite{Nicolas15}.

\vskip 0.05in

In the previous study~\cite{Priezjev15}, molecular dynamics
simulations were carried out to investigate the influence of a local
shear transformation on plastic deformation of a three-dimensional
model glass.      The shear transformation was introduced in a
quiescent system via a spherical inclusion that was gradually
strained into an ellipsoid and then converted back into the sphere
during a finite time interval.         It was demonstrated that at
strain amplitudes above a few percent, the structural relaxation of
the material involved localized plastic events that were identified
using the cage detection algorithm~\cite{BiroliPRL09}.       The
spatial distribution of clusters of cage jumps and their radial
density profiles were studied for various damping conditions and
durations of the shear event.       Interestingly, it was found that
the density profiles of cage jumps are well described by a universal
function multiplied by a factor that depends on the friction
coefficient and the shear transformation time
scale~\cite{Priezjev15}.     It remained unclear, however, how this
factor varies with the strain amplitude and whether it can be
expressed as a function of a single variable.

\vskip 0.05in

In this paper, the plastic response of the amorphous material to a
reversible shear transformation is examined over a wide range of
damping conditions and oscillation time scales.    The analysis of
the density profiles of cage jumps presented in the previous
study~\cite{Priezjev15} is extended further to describe the effect
of strain amplitude on the profile shape and the dependence of the
density maximum on the friction coefficient and oscillation period.
In particular, it is demonstrated that, at sufficiently slow
transformation rates, the peak value of the density profiles is a
function of the ratio of the friction coefficient and the time scale
of the shear transformation and that it strongly depends on the
strain amplitude.

\vskip 0.05in

The rest of the paper is structured as follows.   The details of
molecular dynamics simulation model are described in the next
section.    The analysis of the radial density profiles of cage
jumps as a function of the time scale of the shear event, friction
coefficient, and the strain amplitude is presented in
Sec.\,\ref{sec:Results}.   The conclusions are provided in the final
section.

\section{Molecular dynamics (MD) simulations}
\label{sec:MD_Model}

\begin{figure}[t]
\includegraphics[width=18.cm,angle=0]{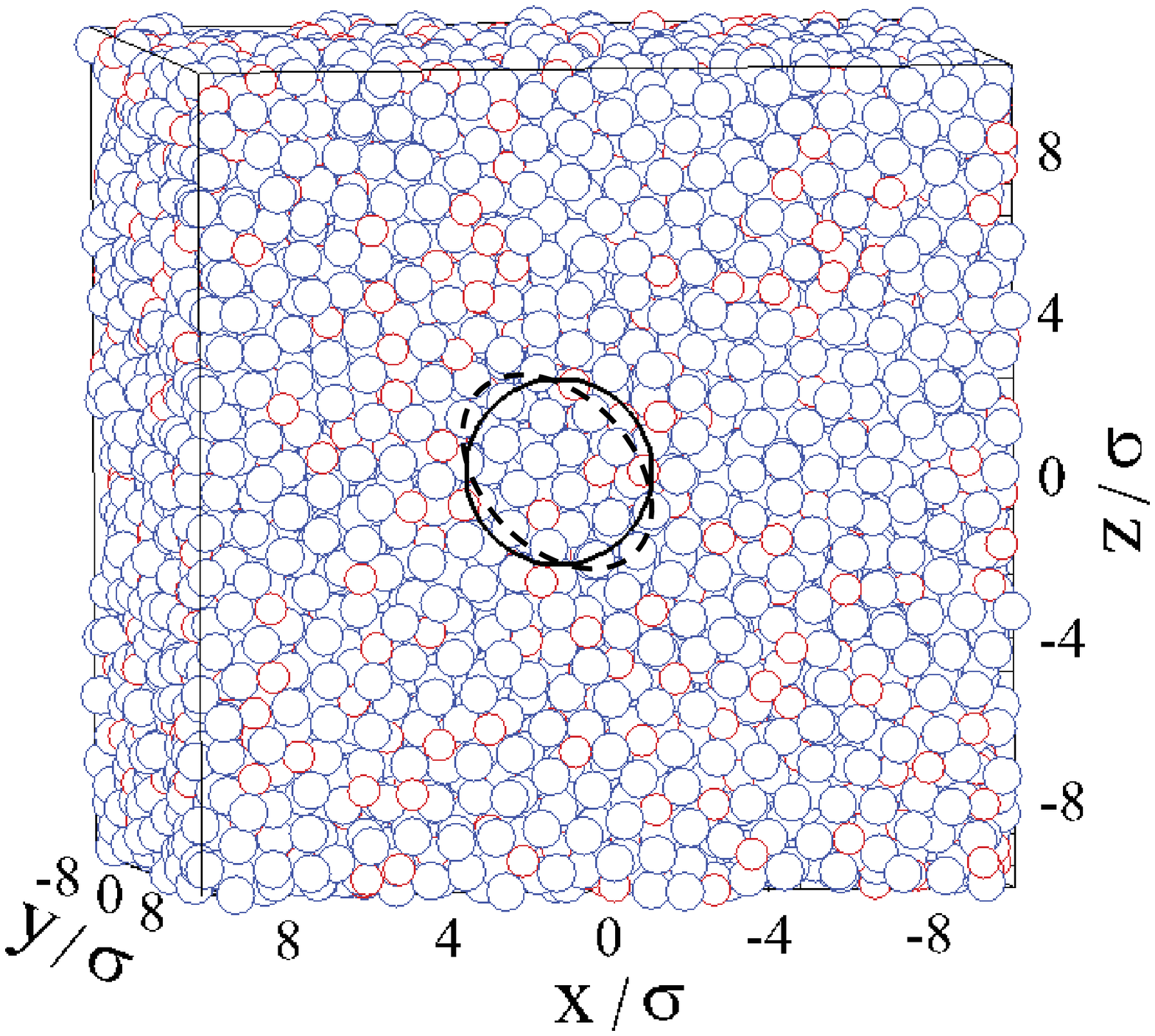}
\caption{(Color online) A snapshot of the instantaneous
configuration of atoms of type A (large blue circles) and type B
(small red circles) in the binary (80:20) LJ mixture.  The spherical
inclusion is located at the center of the periodic cell (black
circle).   The reversible shear transformation is applied to the
inclusion atoms, which are gradually strained into an ellipsoid of
the same volume (dashed ellipse) and then returned to their original
positions. }
\label{fig:snapshot_system}
\end{figure}

The simulated system consists of $N=10\,000$ particles confined in a
three-dimensional cell as shown in Fig.\,\ref{fig:snapshot_system}.
We used a standard model of a glass-forming Lennard-Jones (LJ)
binary mixture introduced by Kob and Andersen~\cite{KobAnd95}. In
this model, any two particles $\alpha,\beta=A,B$ interact through
the Lennard-Jones (LJ) potential
\begin{equation}
V_{\alpha\beta}(r)=4\,\varepsilon_{\alpha\beta}\,\Big[\Big(\frac{\sigma_{\alpha\beta}}{r}\Big)^{12}\!-
\Big(\frac{\sigma_{\alpha\beta}}{r}\Big)^{6}\,\Big],
\label{Eq:LJ_KA}
\end{equation}
with the parameters $\varepsilon_{AA}=1.0$, $\varepsilon_{AB}=1.5$,
$\varepsilon_{BB}=0.5$, $\sigma_{AB}=0.8$, $\sigma_{BB}=0.88$, and
$m_{A}=m_{B}$~\cite{KobAnd95}.   The LJ potential is truncated at
the cutoff radius
$r_{c,\,\alpha\beta}=2.245\,\sigma_{\alpha\beta}$~\cite{Varnik04}.
The units of length, mass and energy are chosen
$\sigma=\sigma_{AA}$, $m=m_{A}$, and $\varepsilon=\varepsilon_{AA}$,
and, consequently, the unit of time is defined
$\tau=\sigma\sqrt{m/\varepsilon}$.   The simulations were performed
at a constant density $\rho=\rho_{A}+\rho_{B}=1.2\,\sigma^{-3}$ and
the linear size of the cubic box is $L=20.27\,\sigma$.   Periodic
boundary conditions were applied along the $\hat{x}$, $\hat{y}$, and
$\hat{z}$ directions.

\vskip 0.05in

The motion of particles is governed by the classical Langevin
dynamics.  For example, the equation of motion in the $\hat{x}$
direction for the $i$-th particle of mass $m$ is given by
\begin{equation}
m\ddot{x}_i + m\Gamma\dot{x}_i = -\sum_{i \neq j} \frac{\partial
V_{ij}}{\partial x_i} + f_i\,,
\label{Eq:Langevin_x}
\end{equation}
where $\Gamma$ is the friction coefficient and $f_i$ is a random
force with zero mean and variance $\langle
f_i(0)f_j(t)\rangle=2mk_BT\Gamma\delta(t)\delta_{ij}$ determined by
the fluctuation-dissipation theorem~\cite{Kubo66}.   The Langevin
temperature is fixed $T=10^{-2}\,\varepsilon/k_B$, where $k_B$ is
the Boltzmann constant.    The equations of motion were integrated
numerically using the fifth-order Gear predictor-corrector
algorithm~\cite{Allen87} with a time step $\triangle
t_{MD}=0.005\,\tau$.     Different realizations of disorder were
prepared by quenching the system from the temperature
$1.1\,\varepsilon/k_B$, which is well above
$T_g\approx0.45\,\varepsilon/k_B$~\cite{KobAnd95}, to the final
temperature $T=10^{-2}\,\varepsilon/k_B$ with the rate of
$10^{-5}\,\varepsilon/k_B\tau$.

\vskip 0.05in


We next describe the deformation protocol for the reversible shear
transformation.   The inclusion atoms were identified within a
sphere of radius $r_i=3\,\sigma$, which is located at the center of
the simulation cell (see Fig.\,\ref{fig:snapshot_system}).   The
average number of atoms in the inclusion is about $135$.    First,
the positions of the inclusion atoms were kept fixed while the
system was aged for about $500\,\tau$ at the temperature
$10^{-2}\,\varepsilon/k_B$.    The spherical inclusion was gradually
strained into an ellipsoid and then converted back into the sphere
during the time interval $\tau_i$.    Note that the major axis of
the ellipsoid was oriented along one of the diagonals of the
simulation box, and the volume of the inclusion was kept constant
during the transformation.    In our study, the shear strain is
defined as the ratio of the ellipsoid semi-major axis to the sphere
radius $r_i=3\,\sigma$.    The variation of stain as a function of
time from zero to $\tau_i$ is described by the following equation
\begin{equation}
\epsilon\,(t) = \epsilon_0\,\,\textrm{sin}(\pi t/\tau_i),
\label{Eq:strain}
\end{equation}
where $\epsilon_0$ is the strain amplitude and $\tau_i$ is the time
scale of the shear event.    After the shear transformation, the
positions of inclusion atoms were kept fixed at their original
positions and the system was allowed to equilibrate for the
additional time interval $10^3\,\tau$.    This time interval is
larger than the damping time $1/\Gamma$ for the smallest value of
the friction coefficient $\Gamma=0.01\,\tau^{-1}$ considered in the
present study.    Several test simulations were performed when the
system was equilibrated for a larger time interval of
$2\times10^3\,\tau$ after the shear transformation in order to
verify that the results remain unchanged.     Finally, the average
atom positions were computed before and after the shear
transformation and then analyzed in $768$ independent samples.

\section{Results}
\label{sec:Results}


At mechanical equilibrium, the atomic structure of the model glass
shows no long-range order while each atom remains trapped in a cage
composed of its neighbors during the time scale of the computer
simulation at the studied temperature.     The plastic deformation
of the amorphous material was induced by a reversible shear
transformation of a spherical inclusion and studied at different
damping conditions and time scales of the shear event.    In our
setup, the inclusion atoms were displaced to form an ellipsoid with
the major axis parallel to the $(1,1,1)$ direction (see
Fig.\,\ref{fig:snapshot_system}) in order to reduce the effect of
periodic boundary conditions. It was observed that at sufficiently
small strain amplitudes (below a few percent), the system response
is elastic and all atoms return to their cages after the shear
transformation~\cite{Puosi14,Priezjev15}.

\vskip 0.05in

In the present study, the analysis of particle positions was
performed in the plastic regime when the strain amplitude was varied
in the range $0.2\leqslant \epsilon_0 \leqslant 0.4$.       We find
that at smaller values of $\epsilon_0$, an accurate analysis of
particle displacements requires averaging over larger number of
independent systems, while at larger strain amplitudes, the relative
distance between inclusion atoms becomes comparable to the molecular
diameter, thus creating voids at the surface of the inclusion during
the shear transformation process.     Irreversible rearrangements of
atoms in the material triggered by the shear transformation were
identified using the cage detection algorithm~\cite{BiroliPRL09}.
Visual inspection of snapshots of the simulated system revealed that
cage jumps tend to aggregate into relatively compact clusters, which
are predominantly located near the inclusion where the deformation
of the material during the shear transformation is
larger~\cite{Priezjev15}.

\vskip 0.05in


It was shown in the previous study~\cite{Priezjev15} that the
clusters of cage jumps are approximately power-law distributed with
an exponent that depends on the strain amplitude.       In general,
it is expected that the density of cage jumps will decay away from
the center of the inclusion because the maximum local strain in the
material during the reversible shear transformation decreases as a
function of the radial distance.     For example, the displacement
and strain fields were calculated analytically for a spherical
inclusion that was strained into an ellipsoid in a two-dimensional
plane, while the third direction reminded
neutral~\cite{Procaccia13}.     It was shown that in a stationary
regime, the strain field has a quadrupolar symmetry and it decays as
$1/r^3$ from the center of the inclusion in three
dimensions~\cite{Procaccia13}.

\vskip 0.05in

\begin{figure}[t]
\includegraphics[width=12.cm,angle=0]{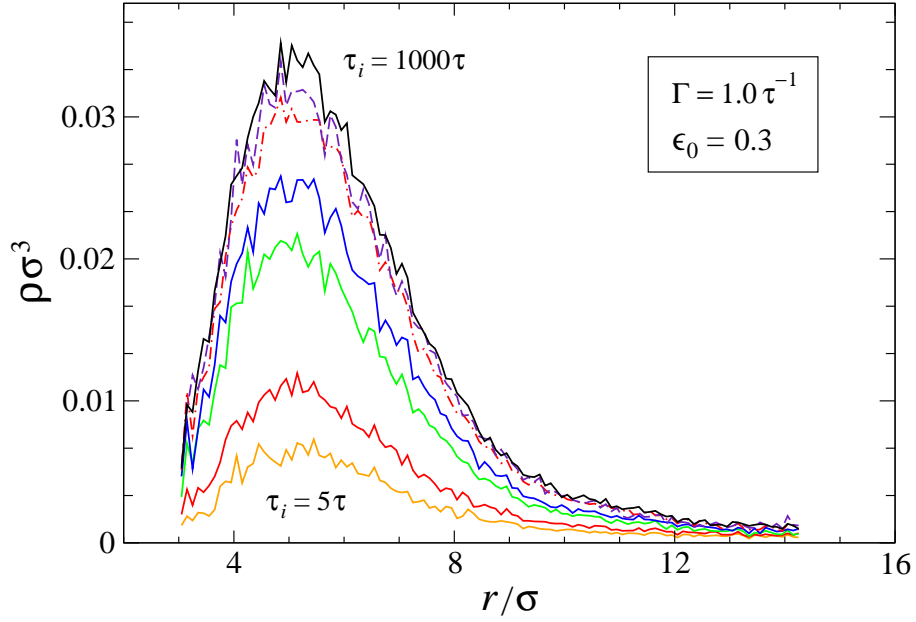}
\caption{(Color online)  Averaged radial density profiles of cage
jumps for the friction coefficient $\Gamma=1.0\,\tau^{-1}$ and the
strain amplitude $\epsilon_0=0.3$.     The time scale of the shear
transformation is $\tau_i/\tau=5$, $10$, $50$, $100$, $300$, $500$,
and $1000$ from bottom to top. }
\label{fig:den_var_time}
\end{figure}

Averaged density profiles of cage jumps as a function of the radial
distance from the center of the inclusion are plotted in
Fig.\,\ref{fig:den_var_time} for the strain amplitude
$\epsilon_0=0.3$ in the regime of intermediate damping
$\Gamma=1.0\,\tau^{-1}$.     Several important features are evident.
First, the density of cage jumps is reduced within about two atomic
diameters from the surface of the inclusion.      This effect
originates from the reversible motion of the inclusion atoms that
effectively form a part of a cage for the neighboring atoms of the
material, thus reducing the probability of their irreversible
displacements~\cite{Priezjev15}.   Second, the density profiles
exhibit a maximum at $r\approx(5-6)\,\sigma$ and then decay with
further increasing radial distance.       It was previously shown
that the rate of decay for $r\gtrsim 6\,\sigma$ correlates well with
the local deformation of the material, which was estimated from the
relative displacement of neighboring particles after the spherical
inclusion was irreversibly strained into an
ellipsoid~\cite{Priezjev15}.

\vskip 0.05in


In general, we find that the average density of cage jumps increases
with increasing time scale of the shear transformation (e.g., see
Fig.\,\ref{fig:den_var_time}).     At small values of
$\tau_i=5\,\tau$ and $10\,\tau$, the time scale of the shear event
is comparable with the time it takes for sound waves to propagate
across the system, and thus the probability of formation of large
clusters is reduced. In contrast, when $\tau_i \gtrsim 50\,\tau$,
the shear stress from the deforming inclusion can induce larger
clusters of cage jumps, which in turn might trigger other
irreversible events in the system.     In addition, as the damping
rate decreases, the effect of inertia becomes more important,
leading to larger avalanches during the shear transformation
process, and, as a result, larger density of cage jumps.     This
trend was also identified in sheared disordered solids in the
athermal quasistatic limit by examining the critical scaling of
avalanches at different damping rates~\cite{Salerno13}.

\vskip 0.05in

\begin{figure}[t]
\includegraphics[width=12.cm,angle=0]{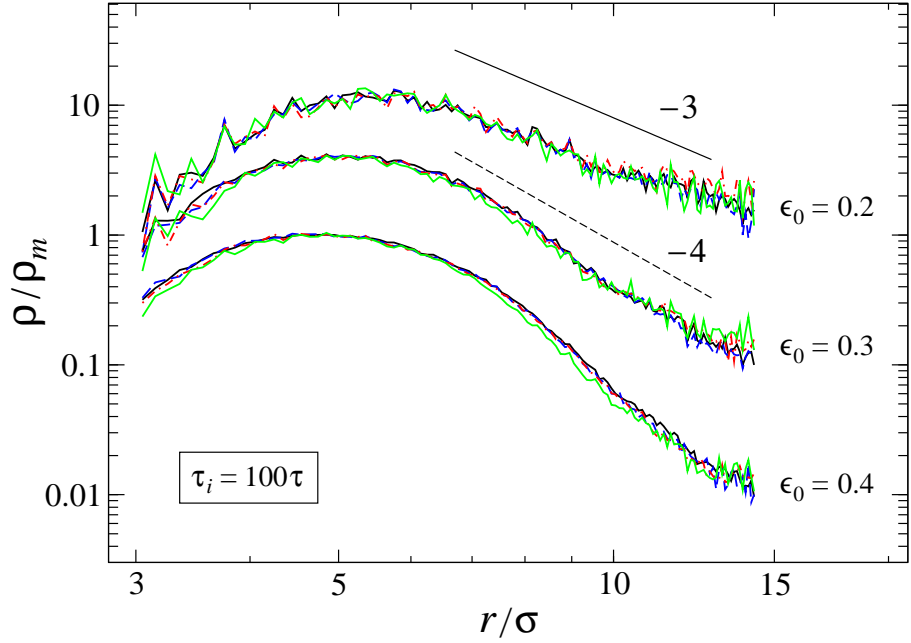}
\caption{(Color online) Log-log plot of the rescaled density
profiles of cage jumps for the strain amplitudes $\epsilon_0=0.2$,
$0.3$, and $0.4$ and the shear transformation time scale
$\tau_i=100\,\tau$.    For each strain amplitude, the friction
coefficient is $\Gamma\tau=0.01$ (solid black curve), $0.1$ (dashed
blue curve), $1.0$ (dash-dotted red curve), and $10$ (solid green
curve).   The data for $\epsilon_0=0.2$ and $0.3$ are displaced
vertically for clarity. The straight solid and dashed lines indicate
slopes $-3$ and $-4$ respectively. }
\label{fig:density_prof_univ}
\end{figure}

It was further noticed in the previous study~\cite{Priezjev15} that
the density profiles of cage jumps for different values of $\tau_i$
and $\Gamma$ can be made to collapse onto a master curve if
$\rho(r)$ is divided by a scaling factor.   In
Fig.\,\ref{fig:density_prof_univ}, we plot the average density
profiles normalized by the corresponding density peak $\rho_m$ for
different values of $\Gamma$ and strain amplitudes $\epsilon_0=0.2$,
$0.3$ and $0.4$.      It is evident that for each value of the
strain amplitude, the rescaled density profiles $\rho(r)/\rho_m$
collapse on the master curves.    Note that the data in
Fig.\,\ref{fig:density_prof_univ} for the two lower values
$\epsilon_0=0.2$ and $0.3$ are shifted for clarity.    It can be
seen that the location of the maximum of $\rho(r)/\rho_m$ and the
slope of decay for $r\gtrsim 6\,\sigma$ depend of the strain
amplitude.   The deviation from the $1/r^3$ dependence, which
describes the decay of the local strain away from an elliptical
inclusion~\cite{Procaccia13}, for larger strain amplitudes
$\epsilon_0=0.3$ and $0.4$ might be due to the finite system size.
Remember that at $r\gtrsim L/2\approx10\,\sigma$ some atoms interact
with their neighbors via periodic boundary conditions but the local
strain is in general not the same across periodic boundaries during
the shear transformation.

\vskip 0.05in

\begin{figure}[t]
\includegraphics[width=12.cm,angle=0]{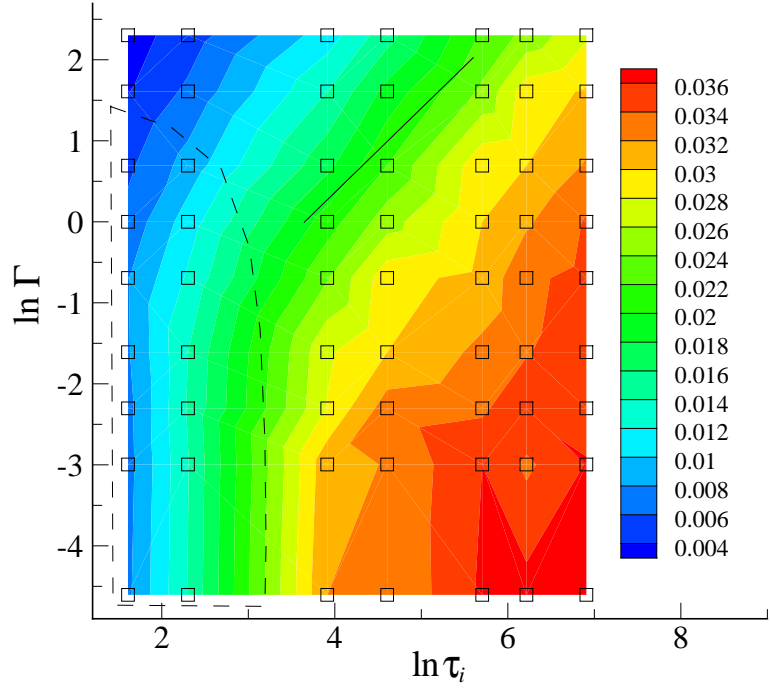}
\caption{(Color online) A contour plot of the density peaks
$\rho_{m}$ (in units of $\sigma^{-3}$) as a function of the friction
coefficient $\Gamma$ and the time scale of the shear event $\tau_i$
for the strain amplitude $\epsilon_0=0.3$. The contour levels are
specified in the legend.   Open square symbols indicate individual
data points.   The straight line with unit slope is shown as a
reference.   The data points within the dashed region were excluded
in the analysis of $\rho_{m}$ presented in
Figs.\,\ref{fig:rhomax_collapse} and \ref{fig:rhomax_eps5} (see text
for details). }
\label{fig:contour_plot}
\end{figure}

In our study, the maximum of the density profiles of cage jumps
$\rho_m$ was estimated in a wide range of parameter values, i.e.,
$0.01\leqslant\Gamma\tau\leqslant10$ and $5\,\tau \leqslant \tau_i
\leqslant 10^3\tau$.    Figure\,\ref{fig:contour_plot} shows a
contour plot of the density peaks $\rho_m$ as a function of the
friction coefficient and the shear transformation time scale for the
strain amplitude $\epsilon_0=0.3$.    It can be seen that the
density landscape is quite complex, but the trends are clear.
Namely, at small $\tau_i$ in the overdamped regime, the deformation
of material during the shear transformation is minimal, and thus the
density of cage jumps is relatively small.   In the opposite limit,
when the shear transformation is very slow and the system dynamics
is underdamped, the deformation of material is largest, facilitating
the formation of large clusters, and the density of cage jumps
saturates to a maximum value of about $0.037\,\sigma^{-3}$ (see
Fig.\,\ref{fig:contour_plot}).


\vskip 0.05in

It can be further observed that the contour lines in
Fig.\,\ref{fig:contour_plot} approximately follow a linear
dependence between the friction coefficient and the time scale of
the shear event (see the straight line with unit slope in
Fig.\,\ref{fig:contour_plot}).    This correlation holds over the
whole range of parameters $\Gamma$ and $\tau_i$, except for the data
points within the dashed region in Fig.\,\ref{fig:contour_plot}.
This, in turn, implies that the ratio $\Gamma/\tau_i$ computed along
a contour level will correspond to the same value of the density
peak. Therefore, it is expected that the data reported outside of
the dashed region in Fig.\,\ref{fig:contour_plot} can be collapsed
onto a master curve if replotted as a function of $\Gamma/\tau_i$.
Another argument for using $\Gamma/\tau_i$ is that the time
dependence of the continuum displacement field after an
instantaneous shear transformation in the overdamped regime is
roughly proportional to the factor $e^{-r^2\,\Gamma/\mu t}$, where
$\mu$ is the shear modulus~\cite{Puosi14}.    Therefore, it follows
that the local displacement field, which can trigger an irreversible
rearrangement of atoms at a distance $r$ from an inclusion, depends
on the ratio $\Gamma/t$.

\vskip 0.05in

\begin{figure}[t]
\includegraphics[width=12.cm,angle=0]{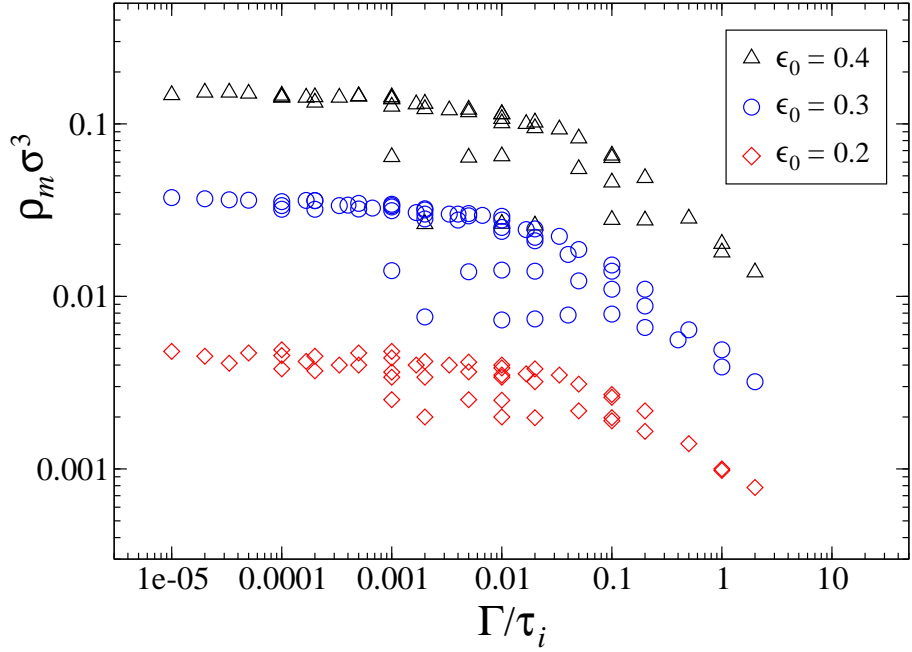}
\caption{(Color online) The variation of the density peak $\rho_{m}$
as a function of the ratio $\Gamma/\tau_i$ for the strain amplitudes
$\epsilon_0=0.2$ ($\diamond$), $0.3$ ($\circ$), and $0.4$
($\triangle$).  The friction coefficient and the time scale of the
shear event vary in the ranges $0.01\leqslant\Gamma\tau\leqslant10$
and $5\,\tau \leqslant \tau_i \leqslant 10^3\tau$, respectively.
Error bars are about the size of the symbols. }
\label{fig:rhomax_all}
\end{figure}
\begin{figure}[t]
\includegraphics[width=12.cm,angle=0]{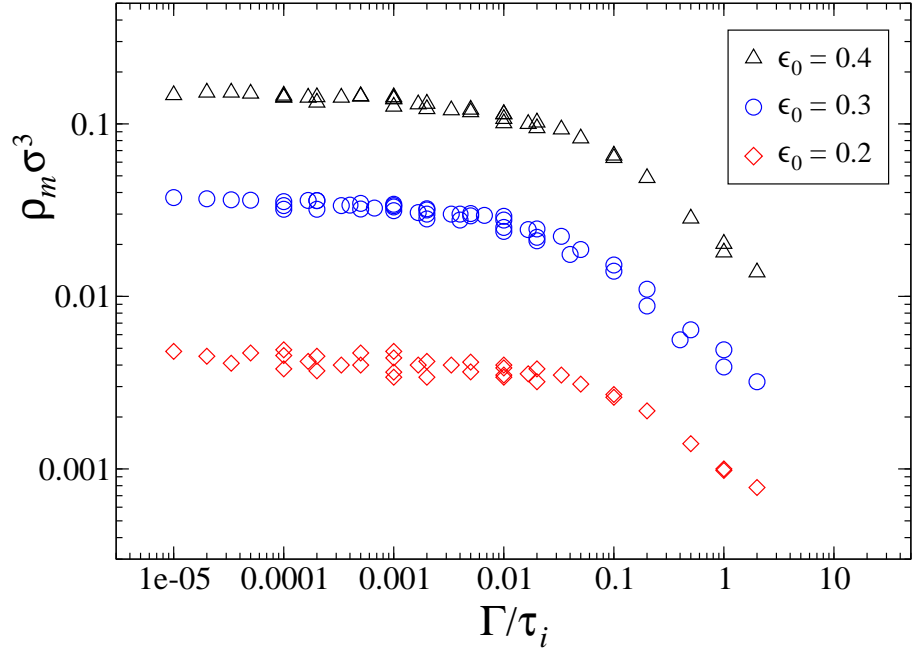}
\caption{(Color online) The variation of the density peak $\rho_{m}$
as a function of the ratio $\Gamma/\tau_i$ for the strain amplitudes
$\epsilon_0=0.2$ ($\diamond$), $0.3$ ($\circ$), and $0.4$
($\triangle$).  The same data as in Fig.\,\ref{fig:rhomax_all}
except for $\Gamma\tau \leqslant 1$ and $\tau_i\leqslant10\,\tau$
(i.e., except for the data points within the dashed region shown in
Fig.\,\ref{fig:contour_plot}).   }
\label{fig:rhomax_collapse}
\end{figure}

The density peaks of cage jumps are first plotted in
Fig.\,\ref{fig:rhomax_all} as a function of the ratio
$\Gamma/\tau_i$ for the strain amplitudes $\epsilon_0=0.2$, $0.3$
and $0.4$ and the same range of parameter values as in
Fig.\,\ref{fig:contour_plot}.    The collapse of the data on three
distinct master curves is satisfactory; however, the data are
somewhat scattered at intermediate values of $\Gamma/\tau_i$.   As
anticipated, the scattered data in Fig.\,\ref{fig:rhomax_all} were
evaluated for the parameters $\Gamma$ and $\tau_i$ within the dashed
region in Fig.\,\ref{fig:contour_plot}.    Without these data
points, the dependence of the density peaks on the ratio
$\Gamma/\tau_i$ is shown in Fig.\,\ref{fig:rhomax_collapse}.   It is
evident that for $\Gamma/\tau_i\lesssim0.01$, the density peak
saturates to a constant value that corresponds to the largest
plastic deformation for each strain amplitude.   In contrast, with
increasing $\Gamma/\tau_i$, the density peak gradually crosses over
to a power-law decay as a function of $\Gamma/\tau_i$. In this
regime, the density of cage jumps is reduced due to either small
time scale $\tau_i$ or large friction coefficient.

\vskip 0.05in

\begin{figure}[t]
\includegraphics[width=12.cm,angle=0]{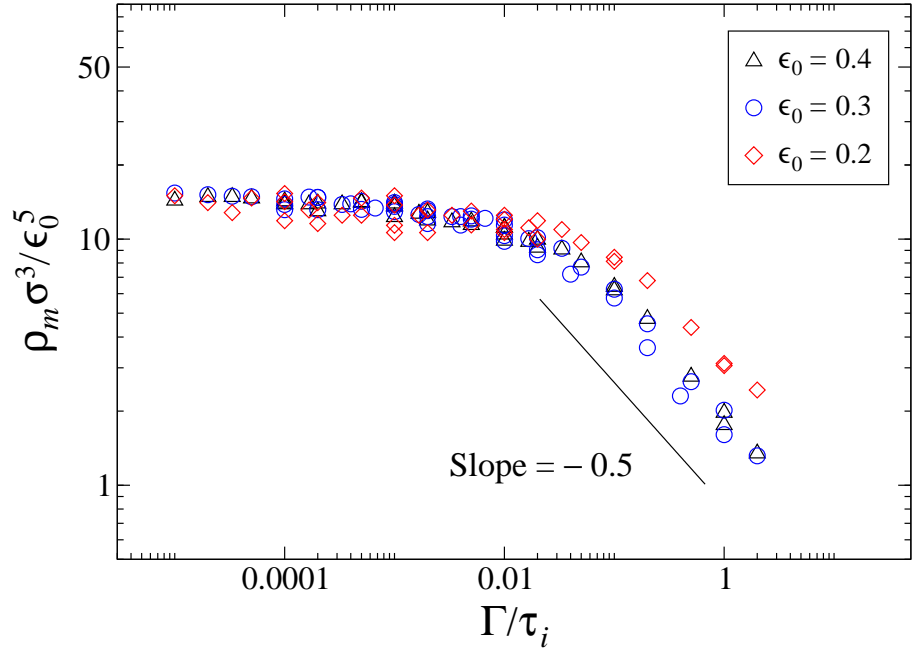}
\caption{(Color online)  Master plot of $\rho_{m}/\epsilon_0^{\,5}$
versus $\Gamma/\tau_i$.     The same data as in
Fig.\,\ref{fig:rhomax_collapse}. The straight line with a slope
$-0.5$ is shown for reference. }
\label{fig:rhomax_eps5}
\end{figure}

It is apparent that the shape of the curves shown in
Fig.\,\ref{fig:rhomax_collapse} is very similar, suggesting that
they might be different by a factor that depends on the strain
amplitude.      Indeed, we found that when the density peaks for
each strain amplitude are divided by $\epsilon_0^{\,5}$, the data
collapse onto a single master curve (see
Fig.\,\ref{fig:rhomax_eps5}).     The resulting master curve extends
over about five orders of magnitude in $\Gamma/\tau_i$.   Notice
that the slope of the decay at $\Gamma/\tau_i \gtrsim 0.01$ depends
slightly on the strain amplitude.    The existence of the plateau in
Fig.\,\ref{fig:rhomax_eps5} implies that the largest value of
$\rho_m$ can be obtained in the limiting case of a very slow shear
transformation for any damping conditions, i.e., when $\Gamma/\tau_i
\rightarrow 0$.     The significance of the value $5$ for the
exponent is at present not clear.     We note, however, that the
exponent was estimated based only on three data points for
$\epsilon_0$ and the critical value of the strain amplitude that
marks the onset of irreversible deformation was not determined in
our study.     The relatively strong dependence of $\rho_m$ on the
strain amplitude might be due the quadrupolar symmetry of the strain
field that can trigger increasingly large clusters of cage jumps
around the inclusion upon increasing strain amplitude.

\section{Conclusions}

In this paper, we have examined the structural relaxation in a
three-dimensional amorphous material induced by a reversible shear
transformation using molecular dynamics simulations.     The
material was deformed by straining a spherical inclusion into an
ellipsoid of the same volume and then converting it back into the
sphere.     We found that at sufficiently large strain amplitude of
the shear transformation, some particles undergo irreversible
displacements that were identified using the cage detection
algorithm.      The density profiles of cage jumps exhibit a
distinct maximum near the surface of the inclusion followed by a
power-law decay as a function of the radial distance.      At a
given strain amplitude, the density profiles are self-similar when
scaled by the density maximum, which in turn depends on the damping
rate and duration of the shear transformation.      Moreover, it was
demonstrated that the data for the peak value of the density
profiles can be collapsed onto a master curve when plotted as a
function of the ratio of the friction coefficient and the
oscillation period.     Overall, these findings indicate that the
density of cage jumps around the inclusion becomes larger in the
cases of weakly damped dynamics or slow shear transformation.

\section*{Acknowledgments}

Financial support from the National Science Foundation
(CBET-1033662) is gratefully acknowledged.  Computational work in
support of this research was performed at Michigan State
University's High Performance Computing Facility and the Ohio
Supercomputer Center.



\bibliographystyle{prsty}

\end{document}